\documentclass[%
 reprint,
 amsmath,amssymb,
aps,
pra
]{revtex4-2}

\usepackage{comment}
\usepackage{graphicx}
\usepackage{dcolumn}
\usepackage[normalem]{ulem}
\usepackage{xcolor}

\usepackage{bm}
\begin{document}

\preprint{APS/123-QED}

\title{Work, Heat and Internal Energy in Open Quantum Systems: A Comparison of Four Approaches from the Autonomous System Framework}

\author{Anja Seegebrecht}
\email{anja.seegebrecht@physik.uni-freiburg.de}
\author{Tanja Schilling}%
\email{tanja.schilling@physik.uni-freiburg.de}

 \affiliation{Institute of Physics, University of Freiburg, Hermann-Herder-Stra\ss e 3, D-79104 Freiburg, Germany.}

\newcommand{\dbar} {\ensuremath{\,\mathchar'26\mkern-12mu d}}
\date{\today}

\begin{abstract}

We compare definitions of the internal energy of an open quantum system and strategies to split the internal energy into work and heat contributions as given by four different approaches from  autonomous system framework. Our discussion focuses on methods that allow for arbitrary environments (not just heat baths) and driving by a quantum mechanical system. As a simple application we consider an atom as the system of interest and an oscillator field mode as the environment. Three different types of coupling are analyzed. We discuss ambiguities in the definitions and highlight differences that appear if one aims at constructing environments that act as pure heat or work reservoirs. Further, we identify different sources of work (e.g.~coherence, correlations, or frequency offset), depending on the underlying framework. Finally, we give arguments to favour the approach based on minimal dissipation.

\end{abstract}

\maketitle


\section{\label{sec:intro}Introduction}

To lay a quantum mechanical foundation of thermodynamics a definition of the terms "internal energy", "work" and "heat" is a prerequisite. The first law of thermodynamics states that the change in internal energy $\Delta U$ is composed of work $W$ and heat $Q$.
In thermodynamics, work is defined as the change of internal energy in the absence of entropic changes, while heat is associated to a change in entropy. At first glance it seems straight-forward to extend this concept to the quantum regime. However, the devil turns out to be in the quantum mechanical details, and over the years a range of disagreeing definitions have been proposed \cite{Alipour_2016,Alipour_2022,Ahmadi_2023,Weimer_2008,Hossein_Nejad_2015,Gemmer_2009,Strasberg_2017,Silva_2021,Beyer_2020,Talkner_2007, Roncaglia_2014, Esposito_2009, Campisi_2011, Mazzola_2013,Rivas_2020}. Ref.~\cite{Dann_2023} provides a recent overview of the differences and relations between single-shot, external, dynamical map, autonomous and semi-classical approaches.

Broadly speaking, the approaches differ by whether they consider the external driving to be exerted by a classical system or by a quantum mechanical one, whether they include the measurement process explicitly in the description, whether the system is already coupled to an infinite bath or a part of the system is integrated out to form an environment (which does not necessarily have to be a bath in the thermodynamic sense), and whether individual systems or statistical ensembles are considered. We cannot give an exhaustive overview of all approaches here and refer the reader to reviews in Refs.~\cite{Dann_2023,Vinjanampathy_2016,Kosloff_2013}.

Our work focuses on methods in the autonomous framework, in which both, the environment and the coupling between the system and the environment have a rather general form, i.e.~we do not restrict the discussion to thermodynamic heat baths, weak coupling or specific initial environmental states. (By the term "thermodynamic heat bath" we understand continuum systems with infinitely many degrees of freedom in a thermal state.) Further, we consider the driving to be exerted by a part of the quantum mechanical system, not by an external classical system. In this regard the methods we discuss here are more general than other approaches. On the other hand, we do not discuss the measurement process, albeit it obviously being necessary for a complete theory. The reason we ignore the measurement process in this article is that we would like to point out some differences between several approaches which have been proposed in the literature. These difference will in general not be removed by a measurement, thus for arguments presented here, the measurement process does not need to be treated.

For a system with a Hamiltonian $H$ characterized by the density matrix $\rho$ the expectation value of the total energy changes according to
\begin{equation}
    \text{d}U= \text{d}\text{Tr}(H\rho) = \text{Tr} (\rho \text{d}H)  + \text{Tr} (H\text{d}\rho) := \dbar W + \dbar Q, \label{eq:conventional_energy_partition}
\end{equation}
where the last two terms are conventionally identified as $\dbar W$ and $\dbar Q$ \cite{Alicki_1979}. In this way work and heat are treated as average quantities and not as distributions.

For a closed system, which cannot exchange energy (or anything else) with the surroundings, energy changes can only be due to a time-dependence of the Hamiltonian. Hence the change of the internal energy is equal to work. 
Usually external driving is imposed by a classical, macroscopic mechanism
\cite{Beyer_2020, Silva_2021} (e.g.~a varying magnetic field).

In contrast, the autonomous framework considers one quantum system as the control or environment for another one. Under certain conditions the environment can be seen either as a work or a heat reservoir, but it could also be a hybrid source. Both components can be treated equivalently such that the assignment of primary system and environment can be inverted.
Tracing out the degrees of freedom of the environment leads to the reduced dynamics of the system of interest.  

The coupling of system and environment causes energy transfer and correlations. Now, the problem arises how to partition the total energy into the internal energy of the primary system and of the generic environment, how to treat energy stored in the correlations, and how to label energy changes as work and heat fluxes. 
Roughly speaking, inclusive and exclusive perspectives can be taken in this context \cite{Jarzynski_2007} as well as perspectives in between the two, i.e.~perspectives in which a part of the interaction Hamiltonian is attributed to the energy operator of the system and in which the environment can contribute to heat and work exchange.

We will analyse and contrast four approaches: the attempt based on a local effective measurement basis (LEMBAS) \cite{Weimer_2008}, the non-local interaction method \cite{Alipour_2016}, the entropy-based formulation derived from spectral decomposition \cite{Alipour_2022, Ahmadi_2023} and the minimal dissipation method \cite{Colla_2021}. The first approach defines an effective system Hamiltonian which is compatible with the initial Hamiltonian and relates heat to entropy changes \cite{Weimer_2008}. 
The second approach aims at a separation of the Hamiltonian, in which the interaction energy is not accessible from the reduced states \cite{Alipour_2016}. The third relies on the spectral decomposition of the system state to relate changes of those eigenvalues which alter the local entropy to heat. And the last approach is based on the minimization of the dissipative part of the equation of motion \cite{Colla_2021}.
 
To compare these methods we consider the spin-oscillator model with different types of interaction as an example.

\section{Methods to define work and heat}
\label{sec:methods}

A closed quantum system, which is isolated in the thermodynamic sense, exchanges neither matter nor energy with the environment. If we describe its state by a density matrix $\rho$, and if the time evolution is generated by a Hamiltonian $H$, it is natural to identify $U = \text{Tr}(H\rho)$
as the internal energy.
The internal energy can only vary in time if the Hamiltonian is time-dependent 
\begin{equation}
    \dot{U} = \text{Tr}(\dot{H}\, \rho) -i\text{Tr}(H [H,\rho]) = \text{Tr}(\dot{H}\, \rho), \label{eq:closed_internal_energy}
\end{equation}
which is usually identified as the work rate \cite{Strasberg_2017}. (Throughout this article we use $\hbar=1$.)

For a genuine quantum mechanical description and in order to consider the most general processes we work in the quantum autonomous framework \cite{Dann_2023}. We start with a composite system that includes the source of driving, such that the total system dynamics are unitary and generated by the static Hamiltonian $H$.
The total state is given by a density operator $\rho$ on the Hilbert space $\mathcal{H}$ and initially prepared in a product state $\rho(0) = \rho_S(0) \otimes \rho_E(0)$, where the index $S$ denotes the system of interest (or primary system) and $E$ an arbitrary environment. In line with this, the Hamiltonian is composed of the bare Hamiltonians $H_S$ and $H_E$ and an interaction term $H_{SE}$. 
The reduced system state is obtained by tracing over the degrees of freedom of the environment, i.e. $\rho_{S} = \text{Tr}_{E}(\rho) $.
This setup defines a completely-positive trace preserving dynamical map $\Phi_t$ which maps the initial system state $\rho_S(0)$ to 
\begin{equation}
    \rho_S(t) := \Phi_t[\rho_S(0)] = \text{Tr}_E(\mathcal{U}(t) \rho_S(0) \otimes \rho_E(0) \mathcal{U}^\dagger (t) ) 
\end{equation}
with unitary $\mathcal{U}(t) = e^{-iHt}$ \cite{Breuer_2002}.
From hereon, unless indicated otherwise, we drop the dependence of the density matrices on time $t$ to keep the notation compact.

The equation of motion for the system can now be written as
\begin{eqnarray}
    \dot{\rho}_S &=& -i \text{Tr}_E [H,\rho] = -i[H_S , \rho_S] -i \text{Tr}_E [H_{SE} , \rho] \\
    &=& -i [H_S+H_S'(t),\rho_S] -i \text{Tr}_E \Big([H_{SE},\chi_{SE}] \Big) \label{eq:EoM_system_state} 
\end{eqnarray}
where we have introduced

\begin{eqnarray}
    H_S'(t) &=& \text{Tr}_E \left(H_{SE} \, (\mathbb{I}_S \otimes \rho_E)\right) \label{eq:correction_Hamiltonian} \\
    \text{and } \chi_{SE} &=& \rho - \rho_S \otimes \rho_E.
\end{eqnarray}

The most common approach identifies $U_S=\text{Tr}_S (H_S\, \rho_S)$ as the internal energy of the system. 
Work $W_S$ done on $S$ is then related to changes in energy due to time-dependence of $H_S$ and heat $Q_S$ is related to the variation of the reduced state \cite{Alicki_1979}. Analogously $U_E = \text{Tr}_E (H_E \,\rho_E)$ can be defined for the environment, but in general one cannot retrieve the internal energy of the composite system by adding $U_E$ and $U_S$ with this choice, because the energy contribution $\text{Tr}(H_{SE}\, \rho)$ is not taken into account. 

Alternatively, a completely inclusive view can be taken by assigning the entire interaction energy to the system of interest  and defining $U_S=\text{Tr}((H_S+H_{SE})\, \rho)$ \cite{Jarzynski_2007}.
This can be inferred if the environment is a heat bath, i.e.~it just exchanges heat~\cite{Esposito_2010, Strasberg_2017}. Variations of the internal energy in $E$ are then identified as heat transferred to or from the system $\dot{Q} = -\dot{U}_E = -\text{Tr}_E (H_E \, \dot{\rho}_E)$. 

With the notion that work contributions only arise due to an explicit time dependence of the Hamiltonian (and $H_E$ being time-independent) the whole work done by or performed on the composite system is attributed to $S$. 
In the limit of weak coupling $\text{Tr}((H_S+H_E)\, \rho) \gg \text{Tr}(H_{SE}\, \rho)$ we obtain the conventional definition given above in Eq.~\eqref{eq:conventional_energy_partition}.

The opposite perspective is taken if the system is coupled to a secondary system which can be labeled as work depository, battery, weight or control system. Work is identified as the negative average energy change of this environment \cite{Skrzypczyk_2014, Dann_2023}: $W=-\text{Tr} (H_E(\rho_E(t)-\rho_E(0))$. 

Generally, these approaches impose particular requirements on the environment or on the coupling. An ideal heat source should remain in a thermal state at all times, while a perfect work reservoir does not build up correlations with the system such that the reduced dynamics of $S$ are unitary. 

The approaches we discuss in the following differ from this idealized view in that they allow the notion of an environment as a hybrid source of work and heat.

Further, the relations above do not take into account that the partition of the total Hamiltonian into system and environment is not unique. The schemes we discuss in the following give distinct motivations for a partition which is applicable to arbitrary setups. We label the effective, rearranged Hamiltonians with superscripts A, B, C and D according to the respective subheadings.

\subsection{Local Effective Measurement Basis}
\label{sec:lembas}

We start with the local effective measurement basis (LEMBAS) approach introduced in Ref.~\cite{Weimer_2008}. The term ``local" indicates that we are only interested in the properties of one component (here either the system $S$ or the environment $E$) of the bipartite system. 
The choice of the measurement basis depends on the context, thus work and heat will be basis dependent. Since in this article we are interested in defining energies in $S$, the eigenbasis of the bare Hamiltonian $H_S$ is a natural choice.

The chosen basis is used as reference for the evaluation of work or heat \cite{Weimer_2008}. The effect of the actual measurement on the dynamics is not considered in this scheme. 
The effective operator is 
\begin{equation}
    H_S^{\rm A}(t) = H_S + H_{S,a}'(t) \label{eq:effective_Hamiltonian_LEMBAS}
\end{equation}
where we have split the correction Eq.~\eqref{eq:correction_Hamiltonian} into $H_S'=H_{S,a}'+H_{S,b}'$, and $H_{S,a}'$  is the part that commutes with $H_S$, while $H_{S,b}'$ is the part that does not commute. In other words, $H_{S,a}'$ is the part of $H_S'$ which is diagonal in the energybasis of $H_S$. The internal energy of the system is determined by $U_S = \text{Tr}(H_S^{\rm A}\rho_S)$. 

Heat flux is then identified as the contribution to the change in internal energy that has an effect on the local von Neumann entropy of the system state $S_S=-\text{Tr}_S(\rho_S \ln \rho_S)$, while work flux is the contribution that does not affect $S_S$.
\begin{eqnarray}
    \dot{Q}_S(t) =&& -i \text{Tr} ([H_S^{\rm A}(t) , H_{SE}] \, \chi_{SE}) \label{eq:LEMBAS_heat} \\
    \dot{W}_S(t) =&& \text{Tr}_S (\dot{H}_{S,a}' (t) \rho_S - i [H_S^{\rm A}(t),H_{S,b}'(t)]\, \rho_S)  \label{eq:LEMBAS_work}
\end{eqnarray}

Note, that the work flux is not simply defined as $\text{Tr}_S(\dot{H}_S^{\rm A}(t)\rho_S)$, but related to unitary dynamics. Further, there can be a change in entropy without a corresponding heat flux \cite{Hossein_Nejad_2015}.

The evaluation of the heat flux still requires knowledge of the entire system state.
It vanishes if the dynamics does not induce bipartite correlations, i.e.\ $\chi_{SE}(t)=0$, or $H_S^{\rm A}$ commutes with $H_{SE}$ \cite{Hossein_Nejad_2015}.  

Since commutation relations clearly govern the energy fluxes it can be useful to break the interaction into $H_{SE} = \sum_k S_k \otimes E_k$ where $S_k$ and $E_k$ are operators that act on the respective subsystem, such that $H_S'(t) = \sum_k S_k \langle E_k\rangle$ (Note that this expectation value is evaluated with respect to the reduced environmental state $\rho_E(t)$: $\langle E_k \rangle_t = \text{Tr}_E (E_k \rho_E(t))$.)  
If both, the system and the environment operators commute with the free Hamiltonians neither heat nor work flows.

In the opposite case ($[S_k,H_S] \neq 0 , \, [E_k,H_E]\neq 0$) the subsystems exchange work and heat with each other. 
Then we have $H_S' = H_{S,b}'$ and thus $H_S^{\rm A}=H_S$, such that
\begin{eqnarray}
    \dot{Q}_S(t) =&& -i \text{Tr} ([H_S , H_{SE}] \,\chi_{SE}) \label{eq:heat_full_commute} \\
    \dot{W}_S(t) =&& - i\text{Tr}_S ( [H_S,H_{S,b}'(t)]\, \rho_S) \label{eq:work_full_commute}
\end{eqnarray}
The fluxes are, in general, unequal~\footnote{only for $[H_S,H_{SE}]=-[H_E,H_{SE}]$ we have $\dot{W}_S+\dot{W}_E = 0, \, \dot{Q}_S+\dot{Q}_E=0$}. 

It is apparent that the initial state of the composite system determines if there are work fluxes. At least one of the subsystems has to feature non-diagonal elements with respect to the energy eigenbasis.

\subsection{Non-local Interaction}
\label{sec:non-local}

To determine how two systems exchange work and heat with each other, for a start the partition of the interaction energy could be considered. In general, correlations $\chi_{SE}$ are built up between the two components and thus part of the total energy will neither be accessible through the system nor the environment state alone. In Ref.~\cite{Alipour_2016} the internal energy of each subsystem is identified as the part which is accessible by arbitrary local measurements.

Instead of looking for a compatible correction to the system Hamiltonian in this approach the idea is to rearrange the Hamiltonian into local parts that are associated with system and environment respectively and a third component that is only available through the composite system state.
This requires $\text{Tr}(H_{SE}^{\rm B} \,(\rho_S\otimes \mathbb{I}_E )) = \text{Tr}(H_{SE}^{\rm B}\, ( \mathbb{I}_S \otimes \rho_E) )= 0$ for the effective interaction Hamiltonian. Defined like this, the interaction energy is neither accessible from the system nor the environment by itself. 
It is given by
\begin{equation}
    H_{SE}^{\rm B}(t) = H_{SE} - H_S'(t) -H_E'(t) + \text{Tr}(H_{SE} \,(\rho_S \otimes \rho_E))
\end{equation}
and accordingly 
\begin{eqnarray}
    H_S^{\rm B}(t) =&& H_S + H_S'(t) - \alpha_S\text{Tr}(H_{SE} \, ( \rho_S \otimes \rho_E) )\label{eq:non-local_effective_H_S}\\
    H_E^{\rm B}(t) =&& H_E + H_E'(t) - \alpha_E\text{Tr}(H_{SE} \, ( \rho_S \otimes \rho_E) ) \label{eq:non-local_effective_H_E}
\end{eqnarray}
where $\alpha_S$ and $\alpha_E$ are free parameters that add up to one. In this framework the internal energy of the entire composite system is made up of the contributions from the subsystems and, additionally, the ``binding energy" $U_\chi = \text{Tr}(H_{SE}^{\rm B}(t)\, \chi_{SE})$. \\
With these choices the work flux can be expressed as
\begin{eqnarray}
    \dot{W}_S(t) &=& \text{Tr} \Big(\dot{H}_S^{\rm B} \, \rho_S\Big) \label{eq:work_non-local}\\
    &=& \text{Tr}\Big(H_{SE}  \,( \alpha_E \rho_S \otimes \dot{\rho}_E - \alpha_S \dot{\rho}_S \otimes \rho_E)\Big) = -\dot{W}_E (t). \nonumber
\end{eqnarray}
The last equality is obtained by treating the environment analogously to the system.
Thus the work fluxes always add up to zero. They clearly depend on the choice of the free parameter and are thus not invariant under gauge transformations \cite{Campisi_2011}. The fact that Hamiltonians $H$ and $H+g(t)\mathbb{I}$ generate the same dynamics but lead to different conclusions for work was pointed out in Ref.~\cite{Vilar_2008} and has caused some debate about the physical significance of the non-local approach. Different setups can imply a specific gauge \cite{Campisi_2011, Alipour_2016}, but a clear prescription for generic situations is not available. 

One possibility would be to remove contributions proportional to the identity which clearly do not influence the dynamics by the effective Hamiltonian. In the non-local approach this would amount to choosing $\alpha_S=0$. 
But this, implicitly, identifies $\alpha_E=1$, such that $\text{Tr}(H_{SE}(\rho_S \otimes \rho_E))\mathbb{I}$ is the last term in $H_S^{\rm B}$, Eq.~\eqref{eq:non-local_effective_H_E}. We see this definition, which combines properties of system and environment, as a drawback of the non-local scheme.

In contrast, the change in heat does not depend on the free parameters
and it has the same structure as Eq.~\eqref{eq:LEMBAS_heat}:
\begin{equation}
    \dot{Q}_S (t) = -i \text{Tr} \Big([H_S^{\rm B}(t),H_{SE}]\, \chi_{SE} \Big) \label{eq:heat_non-local}
\end{equation}
but with a different effective Hamiltonian. 
The heat fluxes add up to the change in binding energy ($\dot{Q}_S+\dot{Q}_E = -\dot{U}_\chi$).

As in the LEMBAS approach, correlations are necessary but not sufficient for a finite heat flux. Or conversely, the environment can also be considered a work source 
if the commutator in Eq.~\eqref{eq:heat_non-local} vanishes.

\subsection{Decomposition}

Recently, a purely entropic motivation for a split of internal energy changes into heat (related to changes in local von Neumann entropy $S_S$) and work (not related to changes in $S_S$) was proposed in Ref.~\cite{Alipour_2022} and Ref.~\cite{Ahmadi_2023}. The authors base their approach on the spectral decomposition of the system state 
\begin{equation}
    \rho_S(t) = \sum_k r_k(t) |r_k(t)\rangle \langle r_k(t)|
\end{equation}
and the observation that the entropy varies if the eigenvalues $r_k$ change: $\dot{S}_S(t) = -\sum_k \dot{r}_k(t) \ln r_k(t)$, but remains constant if just the eigenstates and not the eigenvalues are modified by the dynamics. 
Thus, instead of identifying heat flux as $\text{Tr}_S(H_S \, \dot{\rho}_S) $ only the change in eigenvalues is taken into account:
\begin{eqnarray}
    \dot{Q}_S(t) = \text{Tr}_S \Big(H_S \sum_k\dot{r}_k(t) |r_k(t)\rangle \langle r_k(t)| \Big). \label{eq:decomp_heat_flux}
\end{eqnarray}
The remainder due to the change of eigenstates is attributed to work, thus adding a term to the conventional definition Eq.~\eqref{eq:closed_internal_energy}, i.e.
\begin{eqnarray}
    \dot{W}_S(t) &=& \text{Tr}_S (\dot{H}_S \rho_S) \nonumber\\
    & &+ \sum_k r_k \Big(\langle \dot{r}_k|H_S|r_k\rangle + \langle r_k | H_S|\dot{r}_k \rangle\Big).
\end{eqnarray}

The second term is labeled ``environment-induced dissipative work" caused by counter-diabatic dynamics~\cite{Alipour_2022}. 
Internal energy is conventionally identified with $U_S = \text{Tr}_S(H_S \, \rho_S)$.

These assignments have the advantage that they solely rely on knowledge of the system state and the bare Hamiltonian. However, it seems impractical to decompose the density matrix at each point in time, especially for systems beyond two-level-systems.

The decomposition can be exploited to define a trajectory-based Lindblad-like master equation $\dot{\rho}_S = -i[K_S,\rho_S] + \mathcal{D}_t[\rho_S]$ with 
\begin{eqnarray}
    K_S(t) &=& i\sum_k (|\dot{r}_k\rangle \langle r_k| - \langle r_k| \dot{r}_k\rangle |r_k\rangle \langle r_k|) \\
    \mathcal{D}_t[.] &=& \sum_{k,j} c_{kj} \left[L_{kj} \, . \, L_{kj}^\dagger - \frac{1}{2}\{L_{kj}^\dagger L_{kj}, \, . \, \}\right] \\
    \text{where } L_{kj} &=&  |r_k\rangle \langle r_j|, \qquad c_{kj} = \frac{1-\delta_{r_j 0}}{d} \frac{\dot{r}_k}{r_j}
\end{eqnarray}
and $d$ is the system's dimension \cite{Alipour_2022}. With these identifications the heat flux, Eq.~\eqref{eq:decomp_heat_flux}, can also be expressed as $\dot{Q}_S = \text{Tr}_S(H_S \mathcal{D}_t[\rho_S])$. It vanishes if the eigenvalues of $\rho_E$ stay constant.

Clearly, the effective Hamiltonian $K_S$ obtained in this way, the rates and the Lindblad operators depend on the initial state of $S$. We do not label the effective Hamiltonian $H_S^{\rm C}$, because it is not associated with the internal energy in this approach. 
However, it would seem natural to use $K_S$ as the operator for an inclusive identification of internal energy. Especially, since $H_S$ can be time-dependent in the original works~\cite{Alipour_2022, Ahmadi_2023} and thus, it would be consequent to include the time-dependence, induced by the embedding in an environment, as well. That would be in line with other approaches to define work, heat and internal energy from a master equation perspective~\cite{Colla_2021,Strasberg_2017}, as discussed in the next section~\ref{sec:Min_diss}.

\subsection{Minimal dissipation}
\label{sec:Min_diss}

In Ref.~\cite{Colla_2021} an approach based on the master equation description for the dynamics of an open system is introduced. With the time-convolutionless projection operator technique it is possible to construct an exact, time local master equation for arbitrary coupling strengths and environmental temperatures \cite{Shibata_1977, Chaturvedi_1979}. 

The propagation of an initial system state to some later time $t\geq 0$ can be expressed via the dynamical map $\Phi_t: \rho_S(0) \mapsto \Phi_t[\rho_S(0)] = \text{Tr}_E(\mathcal{U}(t)\, \rho_S(0)\otimes\rho_E(0) \, \mathcal{U}^\dagger(t))$. For varying $t$ we have a family of completely positive and trace-preserving maps \cite{Breuer_2002}. 
The generator of the dynamics is now $\dot{\rho}_S (t) = \mathcal{L}_t[\rho_S] = \dot{\Phi}_t \Phi_t^{-1}[\rho_S(t)]$ if the time-dependence of the map is sufficiently smooth and the inverse exists, which is typically the case \cite{Breuer_2012}. From the requirement of Hermiticity and trace preservation follows that the generator can be rewritten as $\mathcal{L}_t = \mathcal{H}_t + \mathcal{D}_t$ with a Hamiltonian part $\mathcal{H}_t[\, . \,] = -i[K_S(t),\, .\, ]$ and a dissipator with generalized Lindblad structure
\begin{equation}
    \mathcal{D}_t[\, .\, ] = \sum_k \gamma_k \left[L_k \, . \, L_k^\dagger - \frac{1}{2}\{L_k^\dagger L_k, \, . \, \}\right]
\end{equation}
with time dependent rate functions $\{\gamma_k(t)\}$ and operators $L_k(t)$.
This decomposition is not unique, i.e.~for each time $t$ the transformation to Lindblad operators and Hamiltonian
\begin{eqnarray}
    L_k \mapsto L_k &-& \alpha_k \mathbb{I}_S \\
    K_S \mapsto K_S &+& \sum_k \frac{\gamma_k}{2i} (\alpha_k L_k^\dagger  - \alpha_k^\star L_k) + \beta \mathbb{I}_S
\end{eqnarray}
with arbitrary scalar functions $\{\alpha_k(t)\} $ and $\beta(t)$ leaves the generator invariant \cite{Colla_2021}.
However, an unambiguous choice can be made for $\mathcal{D}_t$ if we demand a minimal dissipator which corresponds to traceless Lindblad operators~\footnote{
The dissipator is a Hermiticity and trace preserving superoperator and can be minimized with respect to a norm on this subspace of superoperators. It was shown in \cite{Hayden_2021} that a dissipator with traceless $L_k(t)$ is minimal in this sense.}.  

Once the minimal dissipator is determined the corresponding $K_S$ defines $ H_S^{\rm D}$. Note, that the minimization  does not fix $\beta (t)$ but, because it is irrelevant for the dynamics, it can be set to zero. 

The change of internal energy, work and heat is now given by
\begin{eqnarray}
    \Delta U_S (t) =&& \text{Tr}(H_S^{\rm D}(t)\, \rho_S(t) ) - \text{Tr}(H_S^{\rm D}(0)\, \rho_S(0)) \\
    \delta W_S(t) =&& \int_0^t \text{d} t' \,\text{Tr}(\dot{H}^{\rm D}_S(t')\, \rho_S(t')) \\
    \delta Q_S(t) =&& \int_0^t \text{d} t'\, \text{Tr}(H_S^{\rm D}(t') \, \dot{\rho}_S(t')) \\
    =&& \int_0^t \text{d} t' \, \text{Tr}(H_S^{\rm D}(t') \, \mathcal{D}_t[\rho_S]).
\end{eqnarray}

The advantage of this identification is that it just depends on the system state $\rho_S$ and could, in general, be applied without knowledge about the environment. In addition, it is based on an extremum principle.
However, the useful correspondence between minimal dissipator and traceless Lindblad operators is only established for finite Hilbert spaces. This limitation already causes problems if a single field mode is considered as the system. Even though this can be circumvented by appropriate truncation of the Hilbert space if the dynamics are restricted to a finite subspace, it remains a minor short-coming of an otherwise very convincing approach.

\section{Application to examples}
\label{sec:Application}

To compare these different approaches we study the implications for a simple model and three types of interaction. A two-level system (TLS) or atom is considered as the system of interest and a mode of the radiation field as its environment. 

This is a standard example in quantum optics and it has been considered in various variations in quantum thermodynamics. It is also frequently encountered as an analogue to a steam engine with a working medium (gas, TLS) coupled to a work depository (piston, field mode) \cite{Dann_2023, Tonner_2006}. 
One advantage of this system, and probably the reason for its frequent use as a model, is that it can be treated analytically.
Further, semi-classical analogues are available, that can be exploited for a comparison with semi-classical results 

The free Hamiltonian is $H_0 = H_S+ H_E$ with $H_S = \omega_S \sigma^\dagger$ and $H_E= \sigma + \omega_E a^\dagger a$ where $\omega_S$ and $\omega_E$ denote the frequencies and $\sigma$ and $a$ the lowering operators of atom and field, respectively.

Atom and field are initially uncorrelated, i.e.~they are in a product state $\rho(0) = \rho_S(0)\otimes \rho_E(0)$ with 

\begin{eqnarray}
    \rho_S(0) = \begin{pmatrix} p_e & p_{eg} \\ p_{eg}^\star & p_g \end{pmatrix}
\end{eqnarray}
and the field is in a thermal state, i.e.\ $\rho_E(0) = (1-e^{-\beta \omega_E}) e^{-\beta H_E}$ where $\beta = 1/k_BT$.

For the minimal dissipation approach we have to determine the master equation for the reduced density matrix and identify the effective Hamiltonian and traceless Lindblad operators with corresponding rates. The procedure for this setup is detailed in Appendix~\ref{sec:apx_ME}.

\subsection{Jaynes-Cummings}
\label{sec:JC}

At first we consider the interaction Hamiltonian given by the Jaynes-Cummings model: 
\begin{equation}
    H_{SE}=g( \sigma a^\dagger + \sigma^\dagger a)    
\end{equation}
with interaction parameter $g$. $H_{SE}$ does not commute with the free Hamiltonians of atom and field: $[H_{SE},H_S]= g\omega_S( \sigma a^\dagger-\sigma^\dagger a)$ and $[H_{SE},H_E] = -g\omega_E ( \sigma a^\dagger - \sigma^\dagger a)$.

We obtain the correction Hamiltonian $H_S'=  g(\sigma \langle a^\dagger \rangle + \sigma^\dagger \langle a \rangle) = H_{S,b}'$ which does not commute with $H_S$ (see Eqs.~\eqref{eq:correction_Hamiltonian} and \eqref{eq:effective_Hamiltonian_LEMBAS}).
Consequently, the effective atomic Hamiltonian equals the bare Hamiltonian in 
A: $H_S^{\rm A}=H_S$. In this way $U_S$ corresponds to the conventional identification,  Eq.~\eqref{eq:conventional_energy_partition}. The work and heat fluxes are given by Eq.~\eqref{eq:heat_full_commute} and Eq.~\eqref{eq:work_full_commute}. Only on resonance we obtain zero net work flux $\dot{ W}_S + \dot{ W}_E = 0$ and heat flux $\dot{Q}_S+\dot{Q}_E = 0$, because then $[H_{SE},H_S+H_E]=0$.

\begin{figure}
    \centering
    \includegraphics
    {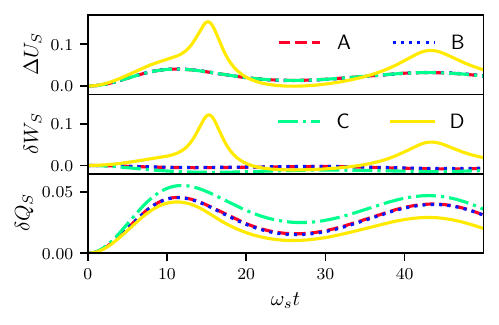}
    \caption{
    Comparison of the integrated energy fluxes for Jaynes-Cummings coupling defined by the methods presented in the previous section~\ref{sec:methods}: A LEMBAS, B non-local (where $\alpha_S=0$ is used), C decomposition, D minimal dissipation for $\omega_E = 0.9 \omega_S, \ k_BT =1.0\omega_S, \ g = 0.1 \omega_S, \ p_e = 0.25, p_{eg}=0.1i$.}
    \label{fig:JC_fluxes}
\end{figure}

In 
B the correction $H_S'$ contributes to the effective local Hamiltonian, provided correlations are built up. In the explored parameter range this results only in a small deviation for the integrated fluxes compared to LEMBAS (compare the lines labelled A and B in Fig.~\ref{fig:JC_fluxes}, where $\alpha_S=0$). For large interaction strengths $g$ and considerable coherence in both, atom and field state, the difference between the quantities increases, because it arises due to $H_{SE}$ and the development of correlations. 

The identification of work and heat in the decomposition scheme (lines labeled C in Fig.~\ref{fig:JC_fluxes}) is qualitatively similar to the previous two approaches. The change of the average internal energy is identical to the LEMBAS picture for this non-commuting interaction. However, the work estimate is in general lower, and the heat estimate higher. By design the decomposition of the state yields a heat flux that is closely related to the local change of entropy (compare the upper panel of Fig.~\ref{fig:IS} with $\delta Q_S$ in \ref{fig:JC_fluxes}: the extrema coincide). 

All three approaches agree that there is no work done by or on the system if the two-level system (and the field) is initially in a diagonal state with respect to the energybasis, i.e.\ $p_{eg}=0$.

The average energy fluxes determined with the minimal dissipation framework are rather different (see lines labeled D in Fig.~\ref{fig:JC_fluxes}). 
Especially, 
the behaviour of internal energy and work can be more "erratic", i.e.~
in comparison to the methods A, B and C we observe peaked features for certain setup parameters. As we approach resonance these features become more prominent. The peaks are not related to the change in entropy or the build-up of mutual information (see Fig. 4) but closely linked to the dynamical map that generates the dynamics. The peaks appear if a singularity is approached, i.e.~a point in time where $\mathcal{L}\rho = \dot{\rho}$ has no unique solution or equivalently, the map $\Phi_t$ is not invertible (see the Appendix~\ref{sec:apx_ME} for details). Whenever the determinant of $F_t$ (the matrix representation of $\Phi_t$) approaches $0$ a peak forms, which does not occur in the other frameworks. A possible interpretation is that a high energetic cost has to be paid to prepare states close to a crossing, where the information about the initial state would be lost. Or conversely a large amount of energy could be extracted by a transformation of the form $U\otimes \mathbb{I}$ on system and environment~\footnote{However, we will not enter the discussion of work extraction schemes here. This would entail questions on how to define ergotropy or how a extraction/measurement protocol should look like and is beyond the scope of this article.}.

Contrary to methods A, B and C, in the minimal dissipation approach a finite energy change associated with work is only obtained if there is a frequency offset of the bare Hamiltonians, i.e.~$\omega_E-\omega_S\neq0$. 
This bias leads to a shift of the populations in the field mode away from a thermal state but remaining diagonal in the eigenenergy basis. Thus one could associate local temperatures to each energy level. The bias offers a resource of work which can be consumed (provided) by $S$ and increase (decreases) the energy stored in the correlations~\cite{Elouard_2023}. 
Initial coherence with respect to the energy eigenbasis of the TLS or the field is not necessary, since the time-dependence of the effective Hamiltonian is determined by the initial state of the environment and the interaction Hamiltonian (see Appendix~\ref{sec:apx_ME}). 
But coherences enhance the exchange of work as well and can be identified as a resource.   

The heat defined by minimal dissipation is always smaller than in the other approaches and thus serves as a lower bound. In the evolution without coherences $\beta \Delta S_S$ and $\delta Q_S$ share the same extrema and on resonance they coincide. 

\subsection{Displaced spin oscillator}
\label{sec:displaced}
Next, we consider another kind of coupling:
\begin{eqnarray}
    H_{SE} = g\sigma_z (a+a^\dagger) 
\end{eqnarray}
which is known as the displaced spin oscillator model. It has already been regarded as an analogue to a steam engine where the oscillator represents the piston that compresses and expands the spin as the ``working gas" \cite{Tonner_2006}. 
The displacement $a+ a^\dagger$ of the oscillator modulates the splitting of the spin's energy levels \cite{Schroeder_2010}. During the interaction the occupations remain constant and mutual information is build up and destroyed again at the field frequency $\omega_E$ even without initial coherence in $\rho_S$ and on resonance. Likewise, the local entropies vary as depicted in Fig.~\ref{fig:IS}. For $p_{eg}=0$ the local von Neumann entropy of the system remains constant  $\Delta S_S=0$ for $p_{eg}=0$ since the .

This interaction is partially commuting since $[\sigma_z, H_S]=0$ and $[a+a^\dagger,H_E]\neq 0$ \cite{Hossein_Nejad_2015}. The correction $H_S'$ for the atomic Hamiltonian is time-dependent and compatible with the bare system Hamiltonian $H_S$ such that with A 
\begin{equation}
    H_S^{\rm A}(t) = \omega_S \sigma^\dagger \sigma + g\sigma_z \text{Tr}_E((a^\dagger + a) \, \rho_E).
\end{equation}

Thus there is no heat flux, because the effective Hamiltonian commutes with the interaction even if correlations are built up between atom and field (see Eq.~\eqref{eq:LEMBAS_heat}). Still, the local entropy $S_S$ oscillates (see Fig.~\ref{fig:IS}). So the entire entropy change is due to entropy production 
\begin{equation}
    \Sigma_S=\Delta S_S- \beta \delta Q_S. \label{eq:entropy_production}
\end{equation}
Note, that the entropy production rate $\dot{\Sigma}_S$ is temporarily negative. This violates the second law of thermodynamics, if it is understood as the positivity of the entropy production rate at all times~\cite{Strasberg_2017}. The violation can be related to non-Markovianity and memory effects in the dynamics~\cite{Colla_2021}.

\begin{figure}
    \centering
    \includegraphics[width=.5\textwidth]{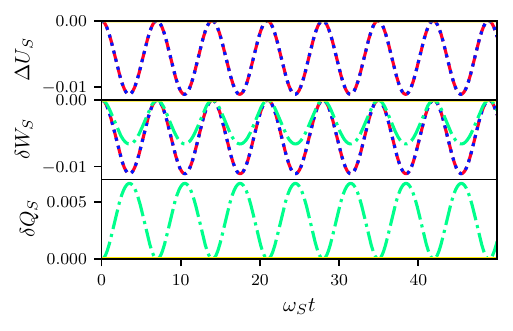}
    \caption{Integrated energy fluxes in the displaced spin oscillator model determined for the same parameter set as in Figure~\ref{fig:JC_fluxes}.  }
    \label{fig:displaced_fluxes}.
\end{figure}

This is also true for B, where $H_S^{\rm B}$ only differs by a scalar (if $\alpha_S=0$, they coincide).
Now, coherence in the initial state is not a prerequisite to obtain a time-dependent correction to $H_S$ and in fact it has no influence on it. Because of the time-dependence both approaches define a work flux and varying internal energy. We observe that $\omega_E$ determines the oscillation frequency of $\Delta U_S, \, \delta W_S$ and the amplitude is given by $-g^2\omega_E$ and $(\alpha_S-1)g^2\omega_E$ respectively (see Fig.~\ref{fig:displaced_fluxes}).

If the instantaneous basis of the system state is employed to infer the thermodynamic energies in C, the results suggest different interpretations.
As internal energy is the expectation value of the bare atomic Hamiltonian $H_S$ and the occupations remain constant during the interaction, we have $\Delta U_S=0$. Although there is no net energy change, there is still an oscillatory exchange of work and heat $\dot{Q}_S =-\dot{W}_S$.

For D the internal energy does not change and there is no work flux because we do not obtain a time-dependent (non-trivial) correction to $H_S$ (see Appendix~\ref{sec:apx_ME}). So in this formulation, it is not sufficient to have coherences or a frequency-offset for a work-like variation of $U_S$.

Although the dynamics is described with a dissipator with non-vanishing rate, it does not contribute to heat since the relevant Lindblad operator $\sigma_z$ commutes with $H_S$ such that:
\begin{equation}
    \text{Tr} (\sigma^\dagger \sigma\, (\sigma_z \rho_S \sigma_z - \rho_S) = 0. \label{eq:sigmaz}
\end{equation}
So the change of local system entropy $S_S$ is clearly not related to energetic changes and solely due to entropy production.

\subsection{Dispersive spin oscillator}
\label{sec:dispersive}
For the sake of completeness, we also consider the dispersive spin-oscillator, an interaction that is fully commuting with the free Hamiltonian:
\begin{eqnarray}
    H_{SE} = g \sigma_z a^\dagger a.
\end{eqnarray}

This coupling leaves the occupations of the atomic levels unchanged and only influences the $S_S$ but not $S_E$.

Since $\sigma_z$ and $a^\dagger a$ commute with $H_S$ and $H_E$ the corrections $H_{S}'(t)=g\sigma_z\langle a^\dagger a \rangle, \, H_{E}'=g\langle \sigma_z\rangle a^\dagger a$ commute with the free Hamiltonian and this results in vanishing energy fluxes in the LEMBAS setting A.

By construction, the non-local approach B does not show any energy fluxes, either. For the heat flux this is obvious because it also depends on the commutator $[H_S^{\rm B},H_{SE}]$ which vanishes, and because the additional scalar in Eq.~\eqref{eq:non-local_effective_H_S} has no effect in the commutator. Correlations between the subsystems are necessary but not sufficient to yield a heat flux. 
The work flux is also zero, because the dynamics only varies the off-diagonal elements in the reduced system states. 

The analysis of the instantaneous eigenbasis C leads to a different result: heat and work are again oscillating and they add up to zero. I.e.~each influx of heat is compensated by an outflux of work and vice-versa. This interpretation suggests, that this setup is not suitable to model a heat or work reservoir.

In contrast, the minimal dissipation approach yields an effective Hamiltonian $H_S^{\rm D}(t)$ that varies in time and leads to a periodic variation of internal energy and work. 
As in the displaced interaction we also obtain a dissipator with a single time-dependent rate and operator $\sigma_z$ (see Eq.~\eqref{eq:sigmaz} and Appendix~\ref{sec:apx_ME}). Thus there is no heat transfer due to these incoherent dynamics. 
Since the entropy is still changing, we can infer that the entropy production rate is directly related to this part of the dynamics.

\begin{figure}
    \centering
    \includegraphics[width=\columnwidth]{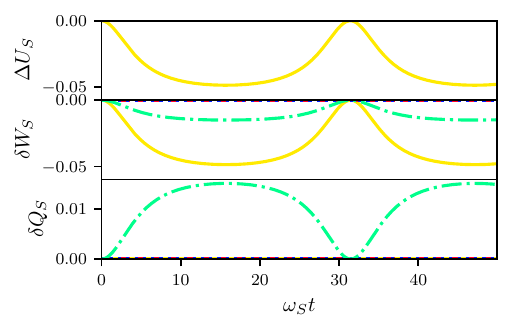}
    \caption{Integrated energy fluxes obtained in the dispersive spin oscillator model. The same parameter set and labels as in Figure~\ref{fig:JC_fluxes} are used.}
    \label{fig:dispersive_fluxes}
\end{figure}

\begin{figure}
    \centering
    \includegraphics[width=\columnwidth]{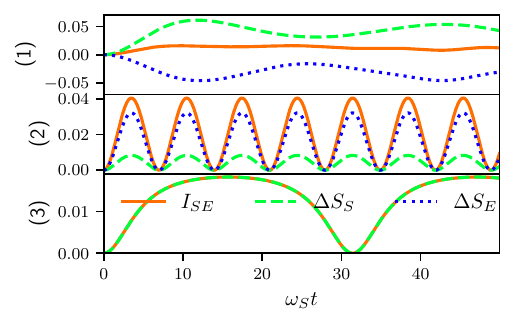}
    \caption{Change of local entropy $\Delta S_S$ and mutual information $I_{SE}=S(\rho)-S_S(\rho_S)-S_E(\rho_e))$ determined with (1) JC, (2) displaced and (3) dispersive interaction for the same parameter set as in Figure~\ref{fig:JC_fluxes}.}
    \label{fig:IS}
\end{figure}

\section{Conclusion}

We have compared four different approaches to identify work, heat and internal energy in bipartite quantum systems. All four can be applied to a system coupled to an arbitrary environment, which does not need to be at equilibrium with a coupling that does not need to be weak. All methods provide a motivation to find a partition of internal energy into a component that locally has a work effect and one that locally has a heat effect. 
Further, they agree that heat is associated with a change in von Neumann entropy, but still reach different partitions. In infinitely large systems, at weak coupling and close to thermodynamic equilibrium, the von Neumann entropy of the subsystem coincides with the thermodynamic entropy\cite{Schroeder_2010}. Accordingly all identifications of heat (and work) should transition to their thermodynamic counterparts in this limit.

In approaches A (LEMBAS), B (non-local) and D (minimal dissipation) work is related to energy exchanges due to changing parameters of the system Hamiltonian. Nonetheless the effective, time-dependent Hamiltonian that arises due to the interaction is defined differently. The LEMBAS framework (A) gives an operational view on work, by presuming that an observer knows the bare system Hamiltonian and will perform measurements with respect to its eigenbasis. So only the internal energy and work accessible in this way is identified.
We note that B gives an ambiguous definition, which is to some extent problematic. The local effective Hamiltonians of $S$ and $E$ are inferred in tandem such that at least one of them includes a term proportional to $\mathbb(I)$ and varying in time, an arguably unphysical situation. 
Criticism might also be brought forward against C (the decomposition scheme) since the treatment of the time-dependence in the coherent part of the dynamics is inconsistent.

Method D offers a unique allocation~\footnote{if the dynamically irrelevant terms in the effective Hamiltonian proportional to $\mathbb{1}$ are neglected} in terms of system variables and it is widely applicable. Apart from exact time-convolutionless master equations the approach can be used for approximate time-local master equations as well. 
The drawback is the limitation of the correspondence between minimal dissipator and traceless Lindblad operators to finite Hilbert spaces\cite{Hayden_2021}. 

If we consider the ability of the methods to identify when an environment acts as a classical driver, A is especially transparent. The variation of the internal energy is solely due to work if the interaction Hamiltonian commutes with the effective system Hamiltonian or if the composite system remains in a factorized state. We observe this for the partially and fully commuting interaction discussed in the spin-oscillator model (see Sec.~\ref{sec:displaced} and Sec.~\ref{sec:dispersive}). But for the latter we also have $\Delta U_S =0$.
In D the heat contribution vanishes if the rates in the dissipator vanish or the Lindblad operators $L_K$ commute with the effective system Hamiltonian. This is equally found for partially and fully commuting couplings, but now $\Delta U_S\neq 0$ for the dispersive interaction and it vanishes for the other.

Quite differently, the diagonalization scheme C does not render any of the interaction types suitable for just driving. 
To obtain a qualitative comparison of how well an environment performs, a measure for the quality of a work source  as suggested in Ref.~\cite{Schroeder_2010} could be used.

Further, we conclude that the different schemes identify different sources. In A and B coherences are essential, while D also ascertains that the off-set in eigenenergies of system and environment can yield an energy change in the form of work.

In addition, the different notions of heat also lead to different interpretations of entropy production if we consider $\Sigma_S = \Delta S_S - \beta \delta Q_S$. Thus, statements about the second law of thermodynamics depend strongly on the underlying definition of thermodynamic quantities.
A detailed discussion would require a particular definition of effective system temperatures applicable in non-equilibrium situations but this is beyond the scope of this investigation.
Further investigation is required to determine which of the formulations is suited best for a discussion of accessibility in measurements and which part of the allocated work can actually be used.

\begin{acknowledgments}
We thank Heinz-Peter Breuer, Alessandra Colla and Graziano Amati for fruitful discussions and helpful comments.
This project has received support by the DFG funded Research Training Group "Dynamics of Controlled Atomic and Molecular Systems" (RTG 2717).

\end{acknowledgments}

\begin{appendix}
\section{Exact master equation}
\label{sec:apx_ME}

To arrive at an effective Hamiltonian and minimal dissipator we follow the strategy given in Ref.~\cite{Smirne}. 

We can cast the dynamical map $\Phi_t: \rho_S(0) \mapsto \Phi_t[\rho_S(0)] = \rho_S(t) = \text{Tr}_E(\mathcal{U}(t) \rho_S(0) \otimes \rho_E(0) \mathcal{U}^\dagger(t))$ in matrix representation and determine the entries numerically. A convenient operator basis is $\{X_k\}_{k=0,1,2,3} = \frac{1}{\sqrt{2}} \{\mathbb{I}, \sigma_i\}$ such that $\Phi_t[\rho_S] = \sum_{k,l} F_{kl} \text{Tr}_E (X_l^\dagger \rho)X_k$ and matrix elements are determined from
\begin{eqnarray}
    F_{kl}(t) = \text{Tr}_S(X_k^\dagger \text{Tr}_E(\Phi_t[ X_l])).
\end{eqnarray}

Then the generator $\mathcal{L}_t$ for the reduced dynamics can be expressed in matrix form with the structure
\begin{eqnarray}
    L(t) = \dot{F}(t)F^{-1}(t) = \begin{pmatrix}
        0 & 0 & 0 & 0 \\ 0 & A & B & 0 \\ 0 & -B & A & 0 \\ X & 0 & 0 & Y 
    \end{pmatrix}
\end{eqnarray} 
provided the inverse does exist, i.e $\det(F) = |\gamma(t)|^2 (\kappa(t) + \eta(t) -1) \neq =0$.
Recast in operator form 
\begin{eqnarray}
    \mathcal{L}_t[\rho_A] =&& iB(t)[\sigma^\dagger \sigma, \rho_S] \nonumber\\
    && + \frac{1}{2}(X(t)-Y(t)) \left(\sigma^\dagger \rho_S \sigma - \frac{1}{2}\{\sigma \sigma^\dagger, \rho_S\}\right) \nonumber\\
    && - \frac{1}{2}(X(t)+Y(t)) \left(\sigma \rho_S\sigma^\dagger - \frac{1}{2}\{\sigma^\dagger \sigma , \rho_S\}\right) \nonumber \\
    && +\frac{1}{4}(Y(t)- 2 A(t) )\left(\sigma_z\rho_S\sigma_z - \rho_S\right). \label{eq:general_ME}
\end{eqnarray}

This yields the effective Hamiltonian $-B(t)\sigma^\dagger \sigma$ which is time-dependent. The dissipator is already in Lindblad form with rates $\frac{1}{2}(X-Y), \, \frac{1}{2}(X+Y), \, \frac{1}{4}(Y-2A)$ corresponding to Lindblad operators $\sigma, \, \sigma^\dagger, \, \sigma_z$. 
These are already traceless and thus we have established the minimal dissipator which can be used to determine internal energy, work and heat in Sec.~\ref{sec:Min_diss}.

For the Jaynes-Cummings interaction (as shown in \cite{Smirne}) the evolution of the atomic state can be described by rates 
\begin{eqnarray}
    A(t) =&& \text{Re} \frac{\dot{\gamma}(t)}{\gamma(t)}, \qquad B(t) = -\text{Im} \frac{\dot{\gamma}(t)}{\gamma(t)} \nonumber \\
    \frac{1}{2}(X(t)&&-Y(t)) = \frac{(\eta(t)-1)\dot{\kappa}(t)-\kappa(t) \dot{\eta}(t)}{\eta(t) +\kappa(t) -1} \nonumber\\
    -\frac{1}{2}(X(t)&&+Y(t)) = \frac{(\kappa(t)-1)\dot{\eta}(t)-\eta(t) \dot{\kappa}(t)}{\eta(t) + \kappa(t) -1 } \nonumber \\
\end{eqnarray}
where
\begin{eqnarray}
    \gamma(t) =&& \langle C(n,t)C(n+1,t)\rangle_F \nonumber\\
    \eta (t) =&&\ \langle C^\dagger(n,t) C(n,t)\rangle _F \nonumber\\
    \kappa (t) =&& \langle C^\dagger(n+1,t) C(n+1,t) \rangle_F \nonumber \\
    C(n,t) =&& e^{i\Delta t/2} \left(\cos(\Omega(n)\frac{t}{2}) -i\Delta \frac{\sin(\Omega(n)\frac{t}{2})}{\Omega(n)}\right) \nonumber \\
    \Omega(n) =&& \sqrt{\Delta^2 + 4g^2n} \nonumber.
\end{eqnarray}
(if the initial field state is diagonal!). 

For the partially commuting interaction 
$X$ and $Y$ are zero, while $A$ and $B$ are found as before from the expectation value with respect to $\rho_F(0)$:
\begin{eqnarray}
    \gamma (t) &=& \langle ( cos(gx)-i\sin(gx))^2 \rangle_E \\ 
    &=& \exp \left(-2g^2 \coth\left(\frac{\beta \omega_E}{2}\right) \frac{1-\cos(\omega_E t)}{\omega_E^2}\right) \nonumber \\
    x &=& i a \frac{e^{-i\omega_E t }-1}{\omega_E} - i a^\dagger \frac{e^{i\omega_E t}-1}{\omega_E}. \nonumber \\
    B(t) &=&  0 \nonumber \\ 
     A(t)  &=& 2 g^2 \coth\left(\frac{\beta \omega_E}{2}\right) \frac{1-\cos(\omega_E t)}{\omega_E^2}
\end{eqnarray}

For the fully commuting interaction $X$ and $Y$ vanish as well, while $A$ and $B$ can be expressed analogously but with 
\begin{eqnarray}
    \gamma(t) &=& \langle (\cos(gnt)-i\sin(gnt))^2\rangle_E = \frac{e^{\beta \omega_E + 2igt}}{-1+e^{\beta \omega_E + 2igt}} \nonumber \\
    B(t) &=&  -i Im \left(  -\frac{2ig}{e^{\beta \omega_E + 2igt}-1}\right)\nonumber \\
     A(t) &=&  Re \left(  -\frac{2ig}{e^{\beta \omega_E + 2igt}-1}\right) \nonumber
\end{eqnarray}

\end{appendix}


%

\end{document}